# Sustainable Recipes[*]

A Food Recipe Sourcing and Recommendation System to Minimize Food Miles

S. Herrera, Juan C.

Nutrition and Food Studies, New York University, New York, NY, USA, jsh501@nyu.edu

**ABSTRACT**

Sustainable Recipes is a tool that (1) connects food recipes ingredient lists with the closest organic providers to minimize the distance that food travels from farm to food preparation site and (2) recommends recipes given a GPS coordinate to minimize food miles. Sustainable Recipes provides consumers, entrepreneurs, cooking enthusiasts, and restauranteurs in the United States and elsewhere with an easy to use interface to help them (1) connect with organic ingredient producers to source ingredients to produce food recipes minimizing food miles and (2) recommend recipes using locally grown food. The main academic contribution of Sustainable Recipes is to bridge the gap between two streams of literature in data science of food recipes: studies of food recipes and studies of food supply chains. The outcomes of the interphase are (1) a map visualization that highlights the location of the producers that can supply the ingredients for a food recipe along with a ticket consisting of their contact addresses and the food miles used to produce a recipe and (2) a list of recipes that minimize food miles for a given GPS coordinate in which the recipe is going to be produced.

## 1 Introduction

Recent trends in food consumption point towards preferences for organic [1]–[4] and local food [5], [6]. In fact, consumers don't perceive local food as expensive and are willing to pay higher prices for them [7], [8]. These trends have generated new players in the food business who entered the market to cater to consumer's preferences ranging from supermarket chains such as Whole Foods and Trader Joe's to food chains like Sweetgreen in the United States to big food companies that acquired stakes in established businesses who provide organic and/or local food such as Stonyfield Farm or Honest Tea to mom and pop startups or restaurants offering local and organic food. For some of these businesses, obtaining information on local food product supply requires research which can be costly and sometimes not completely efficient as it oftentimes relies in traditional methods such as word mouth or searching on the internet. In this paper we propose a computational approach that facilitates obtaining this information as well as recommending food recipes that are local, as in produced with local ingredients. The goal is to propose an efficient way to connect food recipes to organic food production in a twofold way.

First, the proposed tool is geared towards users who want to procure or cook food in a sustainable and environmentally responsible manner through minimizing food miles which will lead to lower carbon emissions, energy savings, time spent by their workers, and lower use of gasoline or electricity to transport food. The end user of this application will be an individual or a food company looking to source ingredients for a recipe of their choice.

Second, this tool seeks to recommend recipes to individuals or food companies keen to minimize the food miles that each ingredient in their recipes will have to travel to the place in which the given recipe is going to be produced. For this reason, the application will favor locally produced ingredients in its recipe recommendation algorithm.

The literature on recipe recommender systems has specialized in two areas. First, studies of food recipes. In particular, these studies look at ingredient combinations to under-stand the rationale behind ingredient pairings [9]–[13], how recipes develop over time [14]–[16], whether geographical proximity is a determinant of recipe similarity [17]–[19], provide recipe recommendations [20] or ingredient substitutions to modify recipes [21]–[23]. Second, studies of food production and supply chains. In particular the role of clusters and social networks in adopting agricultural technologies and practices [24]–[27], traceability, integrity and resilience [28], [29]among others. Furthermore, computational creativity systems have been proposed with the aims of bettering food recipes based on flavor combinations or novelty [30]–[33] .

The primary objective of this paper is to bridge the gap between these two streams of research. We propose a data science approach that connects food recipes and organic food production. Our approach optimizes organic food procurement by minimizing the distances that food must travel to potential businesses hence connecting food recipes and food producers in an optimal way that contributes towards sustainability.

This paper proposes several contributions in the academic literature and on sustainability. First, we aim to bridge the gap in the literature of food recipes and food supply chain by connecting these two streams of research. We hope that this research inspires other types of research, particularly in the realm of computational creativity systems that aim to create recipes as well as in recipe recommender systems. Second, we expect that this research can be translated into real world applications that inform consumers, entrepreneurs, or existing food businesses so they can obtain information regarding their ingredient inputs in a quick, intuitive, and accurate fashion.

Furthermore, in order to make this platform a viable product for users it will undergo several phases. The first phase to implementation will be to partner with local businesses which are interested in sustainability and minimizing food miles to assess the impact of our proposed tool. Afterwards, we will go back to development to incorporate the feedback and consequently launch the tool to be used by the general public. This will make the tool useful for anyone with an internet connection and will help us improve it as in this phase the datasets that power it aren't completely standardized, hence, by allowing users to improve them and provide recommendations we expect that the tool will continually improve to accommodate for the user's needs. Third, we expect that food recipe websites see value in this work so they can include an API to our tool to recommend recipes that minimize food miles and thus provide their website users with more sustainable choices that minimize their carbon footprint.

## 2 Methods

The aim of Sustainable Recipes is to minimize the distance connecting organic products and sites where recipes are produced. We use network science approaches to compute the shortest path in a graph. Since the aim of this tool is bifold, on one hand to connect ingredients to a given recipe and on the other hand, to recommend recipes our approach tackles both problems finding optimal outcomes.

### 2.1 Datasets

The datasets we used are three real world datasets containing information on organic production in the USA, food recipes that are relevant for American consumers, and the locations of Whole Foods, a renowned supermarket chain offering several organic products. Copies of these datasets as well as links to the original locations where they are hosted can be found on the GitHub repository for this project. In what follows we present the information contained in the datasets as well as the transformations that were done to obtain the working datasets.

First, to get information on organic ingredient production we use the United States Department of Agriculture (USDA) Organic Integrity Database which contains information for every organic product recognized by the USDA. This dataset contains an address for each operation (handler and farmer) but not a GPS coordinate. We used the Google maps API to retrieve the GPS coordinates. We then pruned the dataset to contain information of the conterminous 48 US states only. The dataset can be found in the GitHub repository or in its official website: https://organic.ams.usda.gov/integrity/. The dataset was downloaded on June 8 2017 and contains information up to date to 2016. The dataset used for this application containing the GPS coordinates can be obtained here: https://github.com/juancsherrera/Sustainable-Recipes/blob/master/geovector.csv. Hereafter we refer to this dataset as producers dataset.

Second, in order to find information on recipes that are relevant for our study, thus relevant for the American consumers, we used a large dataset containing 56,498 food recipes originally scraped from allrecipes.com by a group of authors [9]. We selected only recipes classified as American since we are focusing on organic products produced in the USA. The resulting dataset contains 35,162 recipes. This dataset has been made publicly available and can be downloaded here: http://yongyeol.com/data/scirep-cuisines-detail.zip For further information on the dataset and the analysis, see The Flavor Network [9]. The final dataset used can be also obtained in the repository for this project here: https://github.com/juancsherrera/Sustainable-Recipes/blob/master/allr_recipes_american.txt. Hereafter we refer to this dataset as recipe dataset.

Third, we used a dataset of Whole Food supermarkets in the United States. This dataset was obtained by scraping the website https://www.wholefoodsmarket.com/stores/list/state on March 22 2018. The corresponding dataset consists of 457 supermarket addresses, however we eliminated 3 supermarkets which were located outside the conterminous 48 USA states; hence the final database consists of 454 supermarkets. These addresses were sent to the Google Maps API in order to obtain the corresponding GPS coordinates. The final dataset, including the GPS coordinates, can be obtained in the GitHub repository containing the code and databases used in this paper: https://github.com/juancsherrera/Sustainable-Recipes/blob/master/wf_geo_public.csv. Hereafter we refer to this dataset as supermarket dataset.

The first two datasets, namely the recipe and producer datasets, are challenging to match as the ingredients on the recipe dataset may not match exactly the products offered by organic farmers and handlers on the producer's dataset. For this reason, we use string matching using the ingredients on the recipe dataset to match parts of the words in the producer's dataset. The goal is to match ingredients that are required to produce a dish sourcing organic ingredient that produced in the conterminous 48 states of the USA. Figure 1 shows the location of all the organic producers and Whole Foods supermarkets. As can be seen, certain areas of the country such as the Northeast and the West Coast concentrate most of the organic production. This has been documented elsewhere [27], with a particular emphasis on peri urban areas whose contours may strengthen biodiversity and local food systems [34] as well as social interactions in local food systems [34].

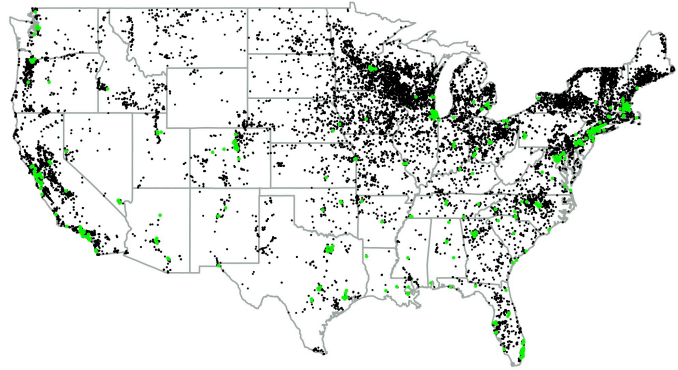

**Fig. 1. Location of all the certified organic producers in the 48 conterminous USA states. Each black dot represents one producer. Total number of producers represented by black nodes: 15,490. Total number of producers represented as green nodes: 454.**

## 2.2 Finding Local Providers

The first problem is to find local providers of ingredients for a given recipe. The method works given two inputs: First, we collect a GPS coordinate which can be selected according to the location of a business, chef, entrepreneur, or home that seeks to produce a recipe. This GPS coordinate represents the location in which the recipe is going to be produced. Second, the recipe producer located in that given coordinate should select a recipe. Afterwards, we collect the information of the ingredients that are needed to produce that recipe. Third, we subset the organic production database to find the ingredients needed to produce the dish. For each ingredient we subset the main database of organic products by matching the words in any of the following three variables: "ci_nopCatName", "ci_nopCategory", or "ci_itemList". These fields contain the products that are recognized certified organic by the United States Department of Agriculture (USDA). Fourth, for the resulting database of producers of each ingredient, we calculate the Euclidean distance between the recipe producer coordinate and each organic producer for a given ingredient. We keep only the edge containing the minimum distance. The rest of the edges are discarded. The process is described in the following algorithm:

---

ALGORITHM 1: Finding Local Providers

Input 1: Recipe production site **GPS** Coordinate
Input 2: Recipe to be produced **R**
Input 3: List of Organic Producers **OP**

For each ingredient **i** in **R**, do:
    Subset database of organic producers **OP** to get only produce of ingredient **i** (string match).
        Return: Subset list **P**
        If **P** is not empty do:
            For each producer **p** in **P**
                Calculate Euclidean distance from **p** to ingredient **i** to GPS
            If distance **p**,**GPS** <= distance **p**,**GIS** keep **p**,**GPS**,
            else drop
            Return: Calculate Food Miles **FM** for recipe: Sum FM for each **GPS, GPSp**
        If P is empty: skip
    end
end
Return list of ingredients, distances, producer name, Euclidean distance, sum Euclidean distances

## 2.3 Local Recipe Recommendation System

The second problem we aim to tackle is to provide recipe recommendations that minimize the distance that the ingredients needed to produce a recipe will travel to a given user. For this, first, we collect a GPS coordinate in which the intended recipe user will be located. Second, for each recipe in our set of recipes we calculate the distance to produce it, or total food miles. For doing so we calculate the distance from the user GPS coordinate to the producer of each one of the ingredients needed to produce the recipe. The result of the algorithm is one recipe from the list of recipes as to ensures that the food miles needed to produce them are minimized. The following algorithm describes the solution to the problem:

---
ALGORITHM 2: Local Recipe Recommendation System
---

Input 1: Recipe production site **GPS** Coordinate
Input 2: List or recipes R
Input 3: List of Organic Producers **OP**
For each ingredient in **R**, do
    For each recipe **r** in **R**
        Subset database of organic producers that produce the ingredient (string match). Generate
        list of producers **P**
            If **P** is not empty do
                For each producer **p** in **P**
                      Calculate Euclidean distance from **p** to ingredient **i** to GPS
                      If distance **p**,GPS0 <= distance **p**,GIS1 keep **p**,GPS1,
                      else drop **p**,GPS1
            If **P** is empty break
            Return: Calculate Food Miles **FM** for recipe: Sum FM for each **GPS, GPSp**
        end
        If total food miles **FM0** recipe <= total food miles **FM1, keep FM1,**
            Else drop **FM1**
    end
end
Return recipe, total food miles

## 3 Results

Given two inputs: a recipe production location and a recipe, we provide a list of producers of ingredients that are needed to make the recipe. This list contains either all the different providers of the ingredient, or the one that is closest to the location that was selected. In this case a random location of a Whole Foods supermarket. This list is accompanied by a visualization showing the location of the producers of each ingredient along with a selection of the nearest producer for each one of the ingredients and a calculation of the total food miles, assuming that each product is brought to the final routing point separately. This can be modified; however, we are assuming that each supplier will deliver to product to the destination separately.

The following example shows our approach for a recipe to be produced in a Whole Foods located in 4520 N. Sepulveda Blvd Sherman Oaks, CA 91403 United States. This recipe contains five ingredients: basil, tomato, wheat, milk, and yeast. The following figures show the supply network selecting the closest location that produces each one of the ingredients. Additionally, we provide a ticket showing the addresses of these producers and the product that they offer along with the total food miles for producing the recipe. Note that this measure is computed only once, that means that if a producer produces two or more ingredients, the distance to that producer is computed only once, not twice or more.

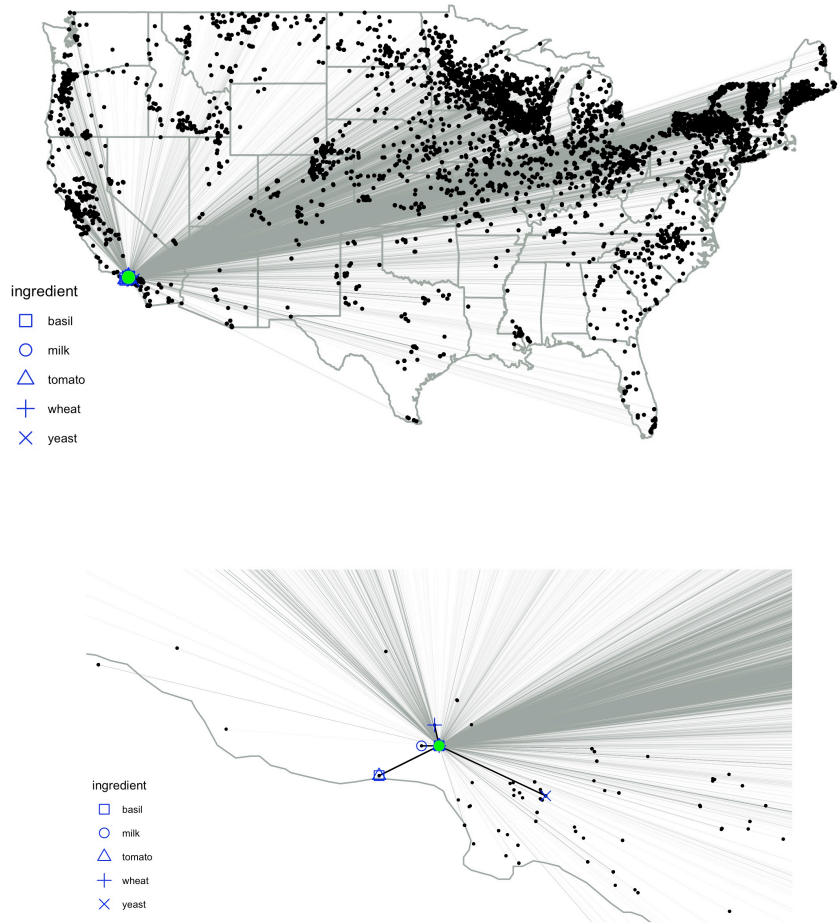

**Fig. 2. Location of all the certified organic producers in the 48 conterminous USA states producing one of the ingredients needed for the recipe; each black node represents one producer. All these producers are linked by gray edges to the randomly selected GPS coordinate representing the location in which the recipe will be produced; green node. The blue nodes show the nearest producer for each one of the products. Since for this recipe all the products are produced very near, we zoomed in.**

**Table 1. Ticket showing the closest supplier for each one of the five ingredients needed to produce a recipe.**

| ingredient | supplier | Product offered by supplier | Distance in miles | Total food miles |
|---|---|---|---|---|
| basil | 22634 Mansie Road Malibu California United States of America the 90265 | Basil | 12.6 | 42.3 |
| tomato | 22634 Mansie Road Malibu California United States of America the 90265 | Tomatoes | 12.6 | 42.3 |
| wheat | 15963 Strathern St. Van Nuys California United States of America the 91406 | Wheat Loaf Bread | 4.6 | 42.3 |
| milk | 60311 Encino Road Rutledge Missouri United States of America the 63563 | Cow's milk | 3.2 | 42.3 |
| yeast | 2429 Yates Ave. Commerce California United States of America the 90040 | Yeast Extract Type Flavor O.S. (AU0179) | 22.0 | 42.3 |

As can be seen in the ticket above, some of the products offered by the supplier do not exactly match with the ingredient. This ticket can further include contact information for each supplier such as the phone number or email address. For example, wheat loaf bread is not exactly wheat. This problem is caused by the nature of the dataset. See the discussion section below.

## 4 Discussion and Conclusions

We have shown a system that tackles the problem of reducing food miles from two sides: by recommending organic food producers that can potentially supply the

## 4.1 Further work

This project which is an a research stage and will be slowly implemented in the future once we have identified local partners, possibly in the Northeast region of the USA has found a way to solve real world problems that will potentially contribute to sustainability by connecting organic food producers, both manufacturers and handlers with recipes either users on a website or businesses that want to produce a recipe. This tool has several shortcomings that have prevented it from implementation. We present the current shortcomings and the future work that we are planning to overcome them.

The main shortcomings of this project arise from three main sources: Methodological constraints and dataset constraints. Here we explain the main causes of the shortcomings and the future work that we are planning to solve them. Additionally, we provide an outline of how we are planning to make this available to the public.

### 4.1.1 Further Work Regarding Methodology:

We do not have data on the size of the operation of the organic products. Hence, in order to make this feasible for a company to source products minimizing the distances we will have to collect data on their potential output. If the potential output does not meet the client's demands, then the node could be reclassified, and the minimization recalculated. The distances between GPS coordinates have been computed as Euclidean distances. This may be problematic in a real-world application; however, this has been done for efficiency and due to lack of credits on the google API. This is an easily solvable problem given some investment money to compute the actual travel distances, nevertheless Euclidean distances provide a good approach. Seasons have not been considered in this analysis. This is a strong shortcoming as we assume that producers can produce and supply food disregard less of the time of the year. A more refined approach should consider this variable in their approach by considering or not considering the node according to the time of the year. In order to overcome these shortcomings, we are planning to partner with local organic producers and restaurants to obtain more information to understand how to overcome these limitations. Furthermore, the recipe recommendation system will incorporate other variables in the future: filters for dietary restrictions in the recipes retrieved, inputs for ingredients that are already on hand (or for which the supply chain can't be modified, etc.), among others.

### 4.1.2 Further work arising from the datasets used:

The Organic integrity database, which is the standard database for organic production is not standardized enough in terms of the actual products that each organic producer offers. Some of the fields in which they list the products are either too broad or too narrow. For example, the field "ci_nopCatName" lists 359 different entries, with such as broad as "Yogurt". The field "ci_nopCategory" is almost empty, showing only 46 entries, and doesn't list yogurt. Lastly, the field "ci_itemList" lists 56,674 different entries and is extremely specific. For example, it lists yogurt, entries such as: "yogurt", "Greek 0% Fat Yogurt - Super Fruits", "Whole Milk Yogurt Pouch - Blueberry", "Squishers Lowfat Yogurt - Strawberry" among others.

Our analysis used word matching in any of the three fields listed before. Hence, for the yogurt example, if any of those fields contained "yogurt", then those producers were matched. This approach provides a quick way to sort through messy data. Nevertheless, in order to provide more accuracy, and make our framework ready for production, it would be necessary to standardize these fields.

Another complication is the name of the recipes, since our recipe dataset does not provide a name for them, we simply provided the recipe recommendations as recipe 1, recipe …, recipe $n$. A production version of this application should include easy to read recipe names that are familiar to users. We have decided to use this dataset because it is well standardized regarding the ingredients and it contains thousands of American recipes which is the geographical focus of our analysis.

#### 4.1.3 Making the findings available:

The aim of this study is to provide an open source platform in which users who are interested in organic products and sustainability can visualize and obtain recommendations to source the organic food products that they need in order to pro-duce a recipe. This tool aims to connect them with the producers. Alternatively, this tool can provide recipe recommendations to users who want to minimize their food miles and support organic producers. As of now the implementation this second part is running very slowly (5 days on a local machine using one GPS coordinate and all the recipe dataset without any radius restriction) even on subsets of the recipe dataset. We expect to improve the running time before releasing the tool.

## 5 Supplemental Materials and reproducibility

All the code (In R) and datasets used in this paper can be found online in the following public GitHub repository: https://github.com/juancsherrera/Sustainable-Recipes